\documentclass[aps,twocolumn]{revtex4}
\usepackage[dvips]{graphics,graphicx}
\usepackage{amscd}
\begin{document}
\title{Dynamics of interacting atoms in driven tilted optical lattices}
\author{Andrey R. Kolovsky}
\affiliation{Kirensky Institute of Physics and Siberian Federal University, 660036 Krasnoyarsk, Russia}
\author{Hans J\"urgen Korsch}
\affiliation{Fachbereich Physik, Technische Universit{\"a}t Kaiserslautern, D-67653 Kaiserslautern, Germany}

\begin{abstract}
The dynamics of cold Bose atoms in driven tilted optical lattices is analyzed focusing on 
destruction of  Wannier-Stark localization and the phenomenon of band collapse. It is argued that 
an understanding of the experimental results requires thorough account for interaction effects. 
These are suppression of the ballistic spreading of atoms for resonant driving (a multiple of 
the driving frequency coincides with the Bloch frequency) and unbounded sub-diffusive spreading of atoms for off-resonant driving.
\end{abstract}
%\pacs{PACS: 05.60.Gg; 05.30.Jp; 03.65.-w\\ Keywords: Bloch oscillations; Destruction of Wannier-Stark localization by interaction induced nonlinearity}

\maketitle

%%%%%%%%%%%%%%%%%%%%%%%%%%%%%%%%%%%%%%
\section{Introduction}
Since the first realization of an atomic Bose-Einstein condensate (BEC) in 1995 much attention is payed to the role of atom-atom interactions in different coherent phenomena of the single-particle quantum mechanics. In particular, restricting ourselves to BECs in optical lattices \cite{Mors06}, papers \cite{Jona03, Mors01} discuss the effect of atom-atom interactions on interband tunneling, papers  \cite{Mors01,Fatt08,Zhen04,Gust08a,09BOBECa} on quasimomentum Bloch oscillations in tilted optical lattices, papers \cite{Bill08,Roat08,Kopi08,Piko08} on the phenomenon of Anderson localization in a disordered or quasiperiodic 1D lattice.

In the present work we study the effect of inter-atomic interactions on BEC's dynamics in driven tilted optical lattices. In recent years driven lattices were intensively studied experimentally with respect to the so-called phenomenon of the Bloch band collapse \cite{Lign07,Sias08,Ecka09}. This term comes from the theoretical prediction for the width of the quasienergy Bloch band, which can take zero values at certain values of the driving amplitude \cite{Dunl86,Dres97}. This effect is also present in driven tilted lattices if the driving frequency is commensurate to the Bloch frequency, defined by the tilt \cite{Sias08}. In this work we revisit the problem of band collapse in driven tilted lattices, focusing on the role of interactions.

The second fundamental problem we address in this paper is the interaction-induced destruction of the Wannier-Stark localization. It is known that for vanishing inter-atomic interactions the eigenfunctions of an atom in a tilted lattice are localized Wannier-Stark states and, hence, any (initially) localized wave-packet remains localized during time evolution. It was argued recently \cite{preprint,Krim09} that, for finite interactions, the time evolution of the wave-packet may be unbounded. It should be stressed from the very beginning that this regime requires weak static forces, not easily accessible in a laboratory experiment. In the opposite limit of a strong static force, the wave-packet dynamics is always bounded, as it has already confirmed experimentally \cite{Gust08}. A new generation of the cited experiment \cite{Gust08} allows to enter the regime of weak static forces by using an additional periodic driving \cite{Hall10,communic}. This opens perspectives for studying the interaction-induced destruction of the Wannier-Stark localization experimentally. %Our theoretical analysis of a BEC dynamics in driven tilted optical lattices precedes these experimental studies.

Our theoretical framework is the following single-particle Hamiltonian
%%1********************************************************
\begin{equation}
\label{1}
H\!=\! -\frac{J}{2}\!\sum_l \big(|l+1\rangle \langle l | + h.c.\big) 
 +d\big[F+F_\omega\cos(\omega t)\big]\!\sum_l |l\rangle l \langle l |,
\end{equation}
where $|l\rangle$ are Wannier states, $J$ the hopping matrix elements, $d$ the lattice period, $F$ the magnitude of a static field, $F_\omega$ and $\omega$ the magnitude and frequency of AC field. Following the detection scheme of the laboratory experiments \cite{Sias08,Hall10,communic}  we are interested in the spacial spreading of an initially localized wave-packet, which we characterize by the square root of the wave-packet second momentum
%2********************************************************
\begin{equation}
\label{2}
\sigma(t)=\Big[\sum_l l^2 P_l(t) -x^2(t) \Big]^{1/2} \;,\quad 
x(t)=\sum_l l\, P_l(t) 
\end{equation}
(here $P_l(t)$ are the occupation probabilities of the lattice sites, $\sum_l P_l(t)=1$). As known, for vanishing interactions the system (\ref{1}) allows an analytical treatment and the quantity $\sigma(t)$ can be calculated exactly for an arbitrary initial state \cite{Moss03,Thom04}. In what follows we borrow the relevant equations for the first and the second wave-packet momentum from \cite{Moss03}, where the model (\ref{1}) was treated by means of dynamical Lie algebras. 

The structure of the paper is as follows. Section~\ref{sec2} is devoted to the dynamics of non-interacting and interacting atoms in a stationary lattice. We recall essentials of the unbounded regime and accomplish studies of Ref.~\cite{Krim09,preprint} by analyzing the rate of the wave-packet spreading in dependence on the strength of atom-atom interactions.  The case of driven lattices is considered in Sec.~\ref{sec3}. It is shown that the problem of BEC dynamics in a driven lattice can be mapped to that in the stationary lattice with properly renormalized static field magnitude and hopping matrix elements. This allows an understanding of the main features of BEC dynamics in driven lattices by referring to the static case of Sec.~\ref{sec2}. The main results of the work are summarized in the concluding Sec.~\ref{sec4}.

%%%%%%%%%%%%%%%%%%%%%%%%%%%%%%%%%%%%
\section{Stationary lattices}
\label{sec2}

\subsection{Single-atom dynamics}
\label{sec-stat-single}

Without driving, $F_\omega=0$, the  eigenfunctions of the Hamiltonian (\ref{1}) are localized Wannier-Stark states
%5********************************************************
\begin{equation}
\label{5}
|m\rangle=\sum_l {\cal J}_{l-m}\big(J/dF\big) |l\rangle \;,
\end{equation}
(here ${\cal J}_\nu(z)$ are Bessel functions of the first kind) and the spectrum is given by the Wannier-Stark ladder with level spacing $dF$. The particle dynamics is a Bloch oscillation (BO) with frequency $\omega_B=dF/\hbar$. It should be mentioned that the character of these oscillations crucially depends on the type of initial conditions. We shall restrict ourselves by considering two limiting cases, 
the case of a completely coherent Gaussian wave-packet of width $\sigma_0\gg 1$,
%3a********************************************
\begin{equation}
\label{3a}
|\psi(t=0)\rangle=\sum_l \sqrt{\rho_l}\, | l\rangle \;,\quad
\rho_l=\frac{1}{\sqrt{2\pi}\,\sigma}\exp\left( -\frac{l^2}{2\sigma_0^2}\right)\,,
\end{equation}
and the case of a completely incoherent wave-packet
%3b********************************************
\begin{equation}
\label{3b}
\hat{\rho}(t=0)=\sum_l \rho_l | l\,\rangle\langle l | \;.
\end{equation}

In the case of coherent initial conditions (\ref{3a}) the system dynamics is the normal BO, where the packet center of gravity 
performs a periodic oscillation with an amplitude given by the Stark localization length ${\cal L}=J/dF$, 
%5a********************************************************
\begin{equation}
\label{5a}
x(t)={\cal L}[1-\cos(\omega_B t)]  \;, \quad \omega_B=dF/\hbar \;,\quad  {\cal L}=J/dF \;.
\end{equation}
During the normal BO the wave-packet width $\sigma(t)$ slightly oscillates, an effect which can be noticed only for a very weak force. This observation helps to understand the limit $F\rightarrow0$, where the wave-packet spreads ballistically with a width $\sigma(t)$ obeying the equation
%4a********************************************************
\begin{equation}
\label{4a}
\begin{CD}
\sigma(t)=\sqrt{\sigma_0^2+(Jt/2\hbar\sigma_0)^2}
@>>{t \rightarrow \infty}>Jt/2\hbar\sigma_0 \;,\quad \sigma_0\gg 1 \;.
\end{CD}
\end{equation}
Note that the rate of ballistic spreading is inversely proportional to the initial width $\sigma_0$.

In the case of incoherent initial conditions (\ref{3b}) the Bloch dynamics corresponds to a so-called breathing mode, where $x(t)=0$ and the wave-packet  width oscillates as 
%5b********************************************************
\begin{equation}
\label{5b}
\sigma(t)=\sqrt{\sigma_0^2+2{\cal L}^2\sin^2(\omega_B t/2)}  \;.
\end{equation}
The $F\rightarrow 0$ limit of this equation gives
%4b********************************************************
\begin{equation}
\label{4b}
\sigma(t)=\sqrt{\sigma_0^2+2(Jt/2\hbar)^2}
\rightarrow Jt/\sqrt{2}\hbar \;,
\end{equation}
where the long time limit is independent of $\sigma_0$. In what follows we shall refer to Eq.~(\ref{4a}) and  Eq.~(\ref{4b}) as slow and fast ballistic regimes, respectively.

%%%%%%%%%%%%%%%%%%%%%%%%%%%%%%%%%
\subsection{Interacting atoms}
\label{sec2b}

To simulate the system dynamics for a finite atom-atom interactions, we solve the discrete nonlinear Schr\"odinger equation,
 %16********************************************************
\begin{equation}
\label{15}
i\hbar\dot{c}_l= -\frac{J}{2}(c_{l+1}+c_{l-1}) +dF  l c_l +g|c_l|^2 c_l \;,
\end{equation}
where $c_l(t)$ is the complex amplitude of a mini BEC associated with $l$th well of an optical lattice and $g$ is the 1D macroscopic interaction constant. Following the structure of the previous subsection we consider both coherent and incoherent initial conditions. According to (\ref{3a}), coherent initial conditions correspond to  
%17********************************************************
\begin{equation}
\label{16}
c_l(t=0)=\sqrt{\rho_l} \exp(i\theta_l) \;,
\end{equation}
with all $\theta_l=0$. To simulate the dynamics for incoherent initial conditions (\ref{3b}) we choose the initial phases $\theta_l$ at random and average the solution of (\ref{15}) over different realizations of random phases $\theta_l$. Typically one needs 10 realizations to get convergence for an integrated characteristic like the wave-packet second momentum $\sigma^2(t)$ and 100 realization to get convergence for the distribution function $P_l(t)=\overline{|c_l^2(t)|}$. 
%###################################
\begin{figure}[b]
\center
\includegraphics[width=8.5cm]{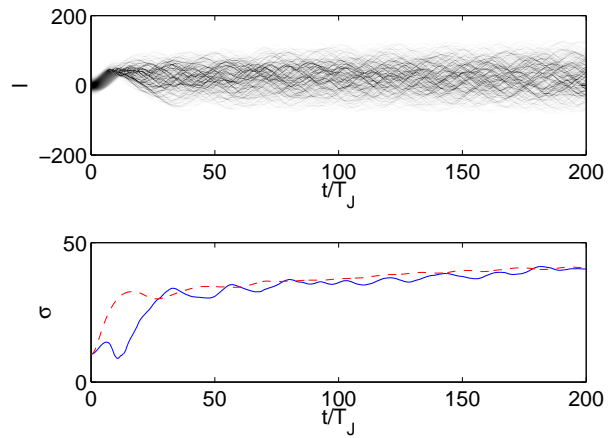}
\caption{Subdiffusive dynamics of a BEC of interacting atoms. Upper panel: Evolution of the atomic density for  coherent initial conditions, $\sigma_0=10$.  Lower panel: $\sigma(t)$ for coherent (solid line) and incoherent (dashed line) initial conditions. Parameters are $J=1$, $dF=0.04$, and interaction constant $g=10$. Time is measured in units of $T_J=2\pi/J$.}
\label{fig3}
\end{figure}

A typical weak-field evolution of the atomic density $P_l(t)$ in the case of coherent initial conditions is shown in the upper panel in Fig.~\ref{fig3}, where time is measured in units of the tunneling period $T_J=2\pi/J$. One can distinguish several stages: The initial short-time dynamics  corresponds to the single-particle BO, where a packet of interacting atoms follows the trajectory (\ref{5a}). This single-particle regime  changes to the regime of dynamical instability (see, e.g., \cite{09BOBECa}) at $t\approx T_B/4$, when the mean quasimomentum crosses the first quarter of the Brillouin zone. As a result the wave packet become scrambled, -- a process which can be viewed as a formation of unstable bright solitons, colliding with each other. During the next stage these solitons `get thermalized' and the time-evolution of the site populations $P_l(t)$ becomes a  random process. These chaotic oscillations of the number of atoms in any given well is a precondition for  the subdiffusive spreading of the atomic cloud predicted in Ref.~\cite{preprint}. 

Before proceeding further we would like to comment on the relation between the mean-field treatment of the system, used throughout the paper,  and the microscopic description based on the many-body Hamiltonian.  With respect to BO this problem was addressed in the recent paper \cite{09BOBECa}. An important conclusion one draws from these studies is that the discussed subdiffusive  dynamics corresponds to an {\em incoherent\/} evolution of the single-particle density matrix. (Typically coherence of an initial BEC state is completely lost after the first 1-3 Bloch cycles.) Thus, when addressing the problem of subdiffusive spreading, one can use an incoherent initial state from the very beginning. In other words, the type of initial conditions affects only the transient short-time dynamics, while the long-time asymptotic dynamics is universal (see lower panel in Fig.~\ref{fig3}). Because the case of a completely incoherent initial packet (\ref{3b}) has certain advantages from the theoretical and numerical points of view, in what follows we shall mainly use these initial conditions. 
%####################################
\begin{figure}[b]
\center
\includegraphics[width=8.5cm]{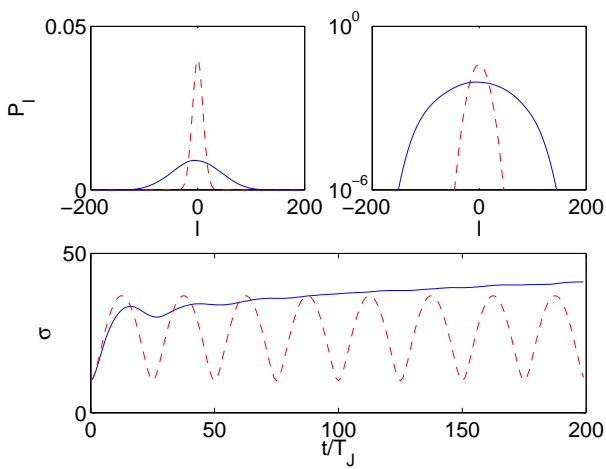}
\caption{Lower panel: $\sigma(t)$  for $g=10$ (solid line) and $g=0$ (dashed line).  Upper panels: Initial (dashed line) and final (solid line) occupation probabilities in linear and logarithmic scale for $g=10$. Incoherent initial state, $J=1$ and $dF=0.04$.}
\label{fig4}
\end{figure}
%####################################
\begin{figure}
\center
\includegraphics[width=8.5cm]{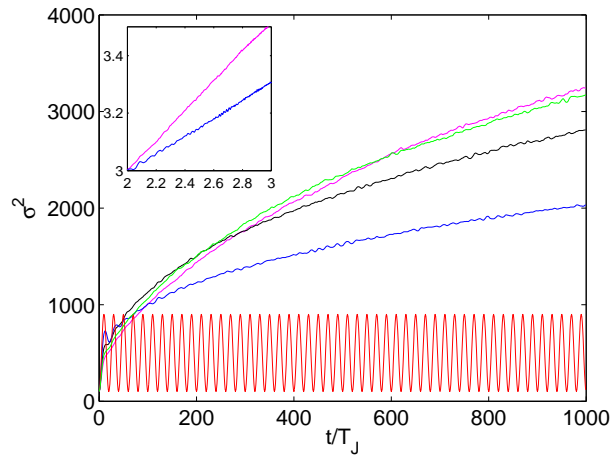}
\caption{The second momentum $\sigma^2(t)$ for the fixed force magnitude $dF=0.05$ and different interaction constant $g=0,10,20,30,40$ (from bottom to top), $J=1$. The inset shows $\sigma^2(t)$ in double logarithmic scale for $g=10$ and $g=40$.}
\label{fig6}
\end{figure}

Let us discuss the characteristic features of the subdiffusive dynamics of interacting atoms in dependence of the system parameters. The dashed and solid lines in the lower panel of Fig.~\ref{fig4} show the behavior of the quantity  (\ref{2}) for $g=0$ and $g=10$, respectively.  It is seen that the initial jump in the wave-packet width is due to the single-particle dynamics, where the maximal packet spreading scales as $1/F$. After this jump the system enters the asymptotic regime, where $\sigma(t)\sim t^{\nu/2}$ with $\nu<1$. The upper panels  in Fig.~\ref{fig4} depict the initial and final distribution of the site populations $P_l(t)$.  Note that this distribution has a well-defined width and, hence, the square root of the second momentum  is a good quantity to characterize the diffusion process \cite{remark1}.  Finally, Fig.~\ref{fig6} shows the behavior of the second momentum $\sigma^2(t)$ for fixed $F$ and different values of the interaction constant $g$.  It is seen that it grows asymptotically as
%6********************************************************
\begin{equation}
\label{6}
\sigma^2(t)\sim t^\nu \;,
\end{equation}
with an exponent $\nu$ depending on $g$.  For a large $g$, where the dynamics of the site populations is fully chaotic, the increment $\nu$ approaches the predicted value $\nu=1/2$ \cite{preprint}.

%%%%%%%%%%%%%%%%%%%%%%%%%%%%%%%%%%%%%%%%%%
\subsection{Unbiased lattices}
\label{sec2c}

For the sake of completeness, this subsection discusses the case $F=0$, which requires a separate consideration. Indeed, for $F\ne0$ and $g\ne0$ the mean-field dynamics of the system is chaotic \cite{remark2}.  Then, after 1-3 Bloch cycles, the system forgets about its initial state and enters the universal asymptotic regime of subdiffusive wave-packet spreading. However, for a vanishing static field the mean-field dynamics is regular and, hence, sensitive to the initial conditions.

It is found that in the case of a wide coherent wave-packet the repulsive interaction {\em  enhances\/} the interaction-free slow expansion (\ref{4a}).  
We also mention that the time-evolution of the distribution function appears to be rather sensitive to the particular shape of the initial wave-packet. For example, for a Thomas-Fermi initial profile, the evolution of $P_l(t)$  differs essentially from that for a Gaussian initial profile even if the wave-packet widths are the same.
%######################
\begin{figure}[t]
\center
\includegraphics[width=8.5cm]{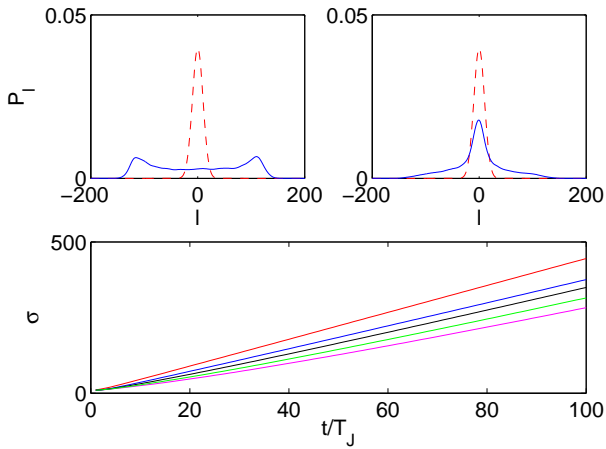}
\caption{ Ballistic spreading of interacting atoms. Lower panel: $\sigma(t)$ for $F=0$, incoherent initial conditions. Values of the interaction constant $g=0,10,20,30,40$ from top to bottom. Upper panels: Occupation probabilities at $t=20\pi$ for $g=0$ (left) and $g=40$ (right), as compared to the initial distribution.}
\label{fig9}
\end{figure}

Unlike the case of coherent initial conditions, a repulsive interaction {\em suppresses\/} the interaction-free fast ballistic regime (\ref{4b}) [see Fig.~\ref{fig9}].  For moderate values of the interaction constant, $g<10$, the characteristic shape and time-evolution of the distribution function $P_l(t)$ resemble those for $g=0$,  which is shown in the upper-left panel in Fig.~\ref{fig9}.  However, for stronger interactions one observes a qualitative deviation from the depicted shape, - the distribution function develops a peak at the origin (compare the upper-right panel). This is a manifestation of the well-known phenomenon of self-trapping,  where the system forms a soliton-like state with an energy outside the Bloch band (see, for example, \cite{Flach04} and references therein). Since the atoms belonging to the soliton state are permanently or temporally (soliton states with a finite live-time) exempt from ballistic spreading, this leads to a decrease in $\sigma(t)$. In what follows, we shall take into account the reduced rate for ballistic spreading of interacting atoms by introducing a suppression coefficient, $C<1$, into Eq.~(\ref{4b}):
%4b********************************************************
\begin{equation}
\label{4c}
\sigma(t;g\ne0)= C(g/J,Jt) \sigma(t;g=0) \;.
%\,\frac{Jt}{\sqrt{2}\,\hbar} \;.
\end{equation}
Although the exact analytical form of $C(g/J,Jt)$ is unknown \cite{remark3}, it is easy to argue that it approaches zero when $g$ is increased or $J$ is decreased. We shall discuss the suppression coefficient in some more details in Sec.~\ref{sec3c} devoted to ballistic spreadings of atoms in driven lattices.

%%%%%%%%%%%%%%%%%%%%%%%%%%%%%%%%%%%%%%%%%%%
%%%%%%%%%%%%%%%%%%%%%%%%%%%%%%%%%%%%%%%%%%%
\section{Driven lattices}
\label{sec3}

\subsection{Single-atom dynamics}
\label{sec3a}

We proceed with driven lattices. Again, our particular interest will be the cases of completely coherent and completely incoherent Gaussian wave-packet. In the former case of coherent packets a generalization of Eq.~(\ref{5a}) for the wave-packet center of mass reads \cite{Moss03}
%18a********************************************************
\begin{equation}
\label{18a}
x(t)=2|\chi(t)|\sin \phi(t)  \;,
\end{equation}
where $|\chi(t)|$ and $\phi(t)$ are the absolute value and the phase of the following complex function
%18b********************************************************
\begin{displaymath}
\nonumber
\chi(t)=J\sum_{n=-\infty}^\infty 
{\cal J}_n\big({\textstyle \frac{dF_\omega}{\hbar\omega}}\big)
{\textstyle \frac{1}{\hbar\Delta\omega_n}}
\exp\big({\textstyle-i\frac{\Delta\omega_n}{2}t}\big)
\,\sin\big({\textstyle\frac{\Delta\omega_n}{2} t}\big) \,,
\end{displaymath}
\begin{equation}
\label{18b}
\Delta\omega_n=\omega_B-n\omega  \;.
\end{equation}
In the latter case of incoherent wave-packets we have $x(t)=0$ and the wave-packet width oscillates as
%18c********************************************************
\begin{equation}
\label{18c}
\sigma(t)=\sqrt{\sigma_0^2 + 2|\chi(t)|^2} \;.
\end{equation}

%#######################################
\begin{figure}[b]
\center
\includegraphics[width=8.5cm]{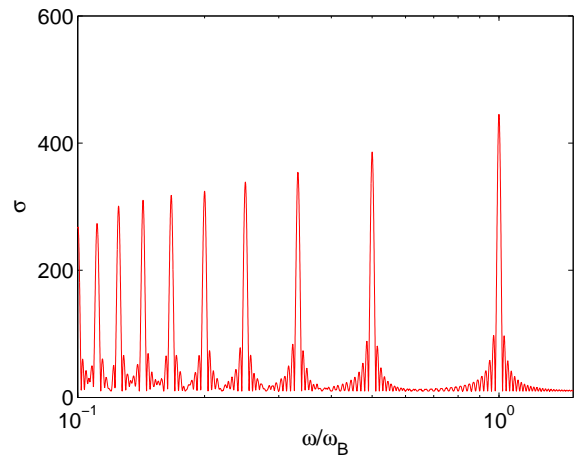}
\caption{Wave-packet width $\sigma(t)$ at $t=200\pi$ as a function of driving frequency  according to Eq.~ (\ref{18c}). Parameters are $J=2$, $dF=0.5$, $F_\omega=1.21F$, and $\sigma_0=10$.}
\label{fig15}
\end{figure}

A characteristic feature of the displayed equations are resonances at integer values of the ratio of the Bloch frequency $\omega_B=dF/\hbar$ to the driving frequency $\omega$ \cite{remark4}. To illustrate such a resonant dynamics of the system,  we depict in Fig.~\ref{fig15} the width $\sigma$ of the incoherent packet at a finite time $t=200\pi$ as a function of the driving frequency. In course of time the hight of each peak grows linearly with $t$,
%19b********************************************************
\begin{equation}
\label{19b}
\sigma(t)= {\cal J}_n\Big({\textstyle\frac{dF_\omega}{\hbar\omega}}\Big)\,\frac{J\,t}{\sqrt{2}\,\hbar} \;,
\end{equation}
while the peak tails show faster and faster oscillations with respect to $\omega$,  with the envelope function approaching
%19********************************************************
\begin{equation}
\label{19}
\sigma_\infty=\frac{ J{\cal J}_n\big(dF_\omega/\hbar\omega\big)}{\hbar|\Delta\omega_n|}  \;.
\end{equation}

Although Eqs.~(\ref{18a}-\ref{18c}) were obtained in a rather formal way, the physics behind these equations is quite simple. To gain a better insight into the near-resonant dynamics we shall use an approach involving the rotating-wave approximation. For simplicity we shall restrict ourselves to the case $\omega\approx\omega_B$ from now on.

It is convenient to present the Hamiltonian (\ref{1}) in the basis of Wannier-Stark states (\ref{5}). We have
%9********************************************************
\begin{equation}
\label{9}
H=dF\sum_m |m\rangle m \langle m| 
+ dF_\omega \cos(\omega t) \sum_{m,m'}|m\rangle V_{m,m'} \langle m'| \;,
\end{equation}
where $V_{m,m'}= \langle m |\left(\sum_l  |l\rangle \,l\, \langle l|\right) |m'\rangle$ are transition matrix elements between different levels of the Wannier-Stark ladder. Due to properties of the Bessel function these matrix elements differ from zero only if $m'=m$ or $m'=m\pm 1$, 
%10********************************************************
\begin{equation}
\label{10}
V_{m,m'}=m\delta_{m',m}+(z/2)\delta_{m',m\pm1} \;,\quad z=J/dF  \;.
\end{equation}
Then, assuming $F_\omega\ll F$ and using the rotating-wave approximation, the quasienergy spectrum of the system is given by the Hamiltonian 
$\widetilde{H} = (dF-\hbar\omega)\sum_m  |m\rangle m \langle m|  + \frac{1}{2}dF_\omega z \sum_m (|m+1\rangle \langle m| + h.c.)$. Finally, introducing an effective static field $d\widetilde{F}=\hbar(\omega_B-\omega)$ and effective tunneling coefficient $\widetilde{J}=J(F_\omega/2F)$ this Hamiltonian takes the form of a Hamiltonian for a stationary lattice,  
 %11********************************************************
\begin{eqnarray}
\label{11a}
\widetilde{H}=\frac{\widetilde{J}}{2} \sum_m (|m+1\rangle \langle m| + h.c.)
+d\widetilde{F} \sum_m |m\rangle m \langle m| \;, \\
\label{11b}
\widetilde{J}=\frac{J}{2}\,\frac{F_\omega}{F} \;,\quad  d\widetilde{F}=\hbar\Delta\omega \;.
\end{eqnarray}
Thus a near resonant driving couples Wannier-Stark states into new `super' Wannier-Stark states with localization length 
%12********************************************************
\begin{equation}
\label{12}
\widetilde{{\cal L}}\approx \widetilde{J}/d\widetilde{F} \sim 1/|\Delta\omega| \;.
\end{equation}
Using this analogy we conclude that a coherent wave-packet in a driven lattice will perform a super BO with a frequency $\Delta\omega$ and an amplitude given in Eq.~(\ref{12}). Of course, one gets the same result directly from Eqs.~(\ref{18a}-\ref{18b}) by keeping in the sum (\ref{18b}) 
only a single term with $n=1$. Moreover, a comparison with these exact expressions indicates that the next after the rotating-wave approximation corresponds to  redefinition of $\widetilde{J}$ as
%13********************************************************
\begin{equation}
\label{13}
\widetilde{J}=J{\cal J}_1\big(F_\omega/F\big)  \;.
\end{equation}
Thus the amplitude of the super BO is a nonlinear function of the driving amplitude, - a phenomenon similar to the phenomenon of band collapse.
%#########################
\begin{figure}
\center
\includegraphics[width=8.5cm]{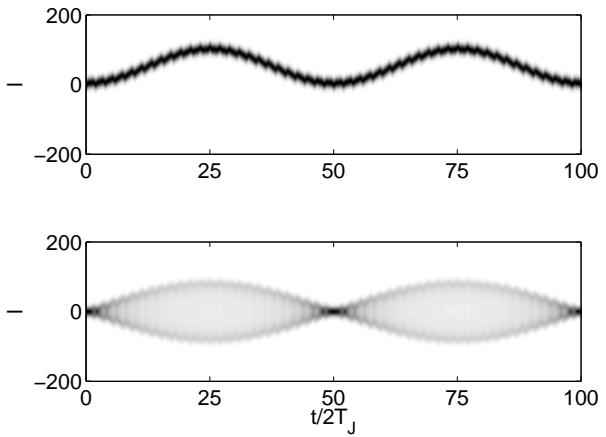}
\caption{Super Bloch oscillations. Numerical simulation of the system dynamics for coherent (upper panel) or incoherent (lower panel) initial states. Parameters are $J=2$, $dF=0.5$, $F_\omega=1.21F$ (hence $\widetilde{J}=1$), and $\hbar\Delta\omega=0.02$.}
\label{fig7}
\end{figure}

As an illustration of the above analysis Fig.~\ref{fig7} shows the wave-packet dynamics for $J=2$, $dF=0.5$, $F_\omega=1.21F$, and off-resonant driving $\Delta\omega=0.02$, which correspond to $\widetilde{J}=1$ and $d\widetilde{F}=0.02$. The packet is seen to oscillate with the usual Bloch frequency $\omega_B=d F/\hbar$ and simultaneously show a slow BO with frequency $\Delta\omega$ and essentially larger amplitude, given in Eq.~(\ref{12}).  We note in passing that such a super BO, depicted in the upper panel, has recently been observed for a BEC of cesium atoms in Ref.~\cite{Hall10}. The breathing mode of super BO, depicted in the lower panel, was observed with not condensed strontium atoms in the 
experiment \cite{Albe09}.

%%%%%%%%%%%%%%%%%%%%%%%%%%%%%%%%%%%%%%%%%
\subsection{Interacting atoms}

\begin{figure}[t]
\center
\includegraphics[width=8.5cm]{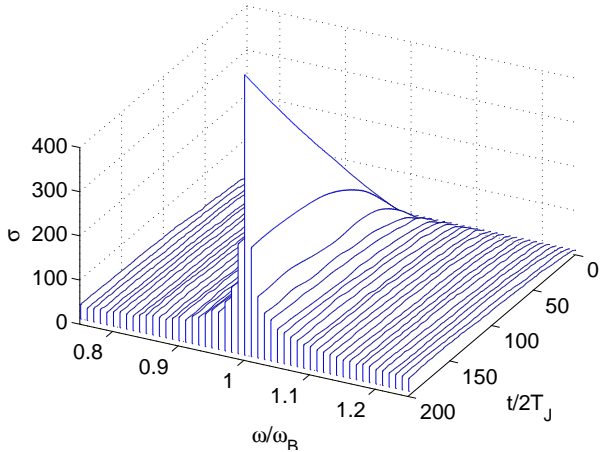}
\caption{Formation of the resonance peak at $\omega=\omega_B$ in the case of interacting atoms, $g=40$. The other parameters are as in Fig.~\ref{fig7} (incoherent initial conditions).}
\label{fig17}
\end{figure}

\begin{figure}
\center
\includegraphics[width=8.5cm]{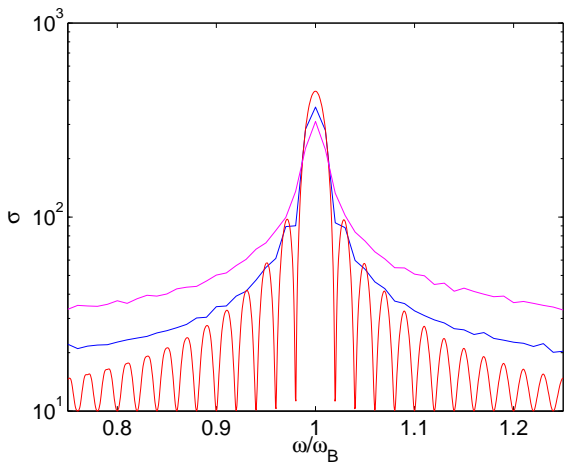}
\caption{The shape of the resonance peak at $t=200\pi$.  The smooth red line corresponds to Eq.~(\ref{18c}) and the two broken lines show results of direct numerical simulations for $g=10$ (blue) and $g=40$ (magenta).} 
\label{fig16}
\end{figure}

For off-resonant driving, the effect of atom-atom interactions on atomic dynamics was found to  be equivalent  to that in a stationary lattice, providing the former lattice is discussed in terms of its effective Hamiltonian (\ref{11a}). Thus, similar to the case of stationary lattices, interactions destroy the super BO after a few super periods $T=h/|\Delta\omega|$. The foremost consequence of the resulting incoherent dynamics is the formation of a {\em smooth\/} resonance peak (see Fig.~\ref{fig17}). Moreover, in course of time the peak shape starts to deviate from (\ref{19}) due to  a slow increase in the wave-packet width (see Fig.~\ref{fig16}). This change in the shape of the resonance peak may serve as an indicator of subdiffusive dynamics.

\begin{figure}
\center
\includegraphics[width=8.5cm]{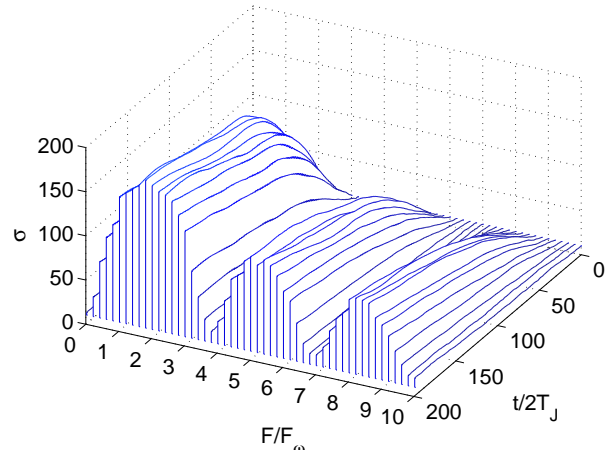}
\caption{Time evolution of the width $\sigma(t)$ for $g=40$ and different $F_\omega$. The other parameters are $J=2$, $dF=0.5$, and $\hbar\Delta\omega=0.02$.}
\label{fig10c}
\end{figure}

It is interesting to study the subdiffusion with respect to the phenomenon of band collapse, described by Eq.~(\ref{13}). For this purpose we simulate the system dynamics for finite detuning $\Delta\omega$, finite interaction constant $g$, and different driving amplitudes $F_\omega$. It is seen in Fig.~\ref{fig10c} that at zeros of the Bessel function, where $\widetilde{J}=0$, the subdiffusive spreading is suppressed almost completely. 
%#####################
%######################
%######################

%%%%%%%%%%%%%%%%%%%%%%%%%%%%%%%%%%%%
\subsection{Ballistic regime}
\label{sec3c}

Finally we discuss the case of an exact resonance. For $\omega=\omega_B$ the AC field couples localized Wannier-Stark states into extended states and the wave-packet spreading is ballistic. However, according to results of Sec.~\ref{sec2c}, this ballistic regime may be different for different types of initial conditions. We have found that the case of incoherent initial conditions (random phases of the complex amplitudes) is well captured by the effective model  (\ref{11a}) but the effective model fails to describe the dynamics of the original system for a coherent initial state, which is actually realized in a laboratory experiment. The reason for this is that the effective model focuses on super BO and overlooks ordinary BO, which 
appears to be important specifically at exact resonance. In what follows we analyze this situation in some more detail.

\begin{figure}
\center
\includegraphics[width=8.5cm]{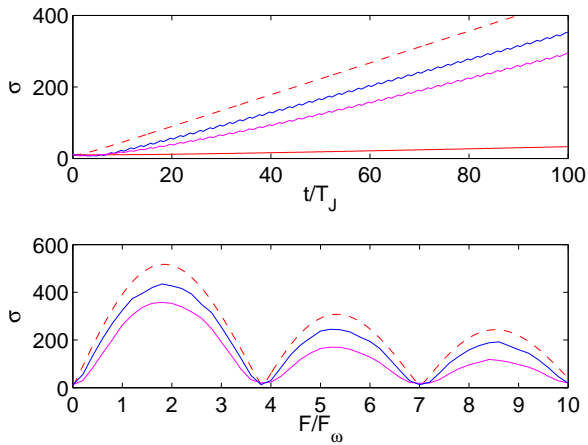}
\caption{Ballistic spreading for $\Delta\omega=0$. Upper panel: The width $\sigma(t)$ for $g=0$ (red line), $g=10$ (blue), and $g=40$ (magenta) and coherent initial conditions. The dashed red line indicates the interaction-free fast ballistic regime for incoherent initial conditions. Lower panel: The width $\sigma(t)$ at $t=200\pi$. The broken lines show results of direct numerical simulations for $g=10$ (blue), and $g=40$ (magenta) for incoherent initial conditions. The dashed red line corresponds to Eq.~ (\ref{4b}) with $J$ substituted by $\widetilde{J}$.}
\label{fig18}
\end{figure}

Let us assume for the moment a non-interacting case. For $g=0$ and $\omega=\omega_B$ the coherent wave packet performs a normal BO with slowly increasing packet width, which is well approximated by Eq.~(\ref{4a}) for slow ballistic spreading (solid red line in the upper panel 
in Fig.~\ref{fig18}). However, the ordinary BO may be also dynamically unstable if $g\ne 0$.  This dynamical instability leads to an exponentially fast randomization of the relative phases of the complex amplitudes $c_l(t)$. As soon as the phases become randomized, the slow ballistic regime (\ref{4a}) changes to the fast one (see the upper panel in Fig.~\ref{fig18}).  Thus the fast ballistic spreading seems to be a generic case in driven lattices, independent of the type of initial conditions \cite{remark5}.

Using results of Sec.~\ref{sec2c}, the fast ballistic spreading is given by Eq.~(\ref{4c}), where one should substitute the hopping matrix element $J$ by the effective hopping matrix element (\ref{13}). 
%13********************************************************
%\begin{equation}
%\label{4d}
%\sigma(t)\approx C\big(\widetilde{J}/g\big)\,\frac{\widetilde{J}\ t}{\sqrt{2}\,\hbar} \;,\quad
%\widetilde{J}=J\,{\cal J}_1(F_\omega/F)  \;.
%\end{equation}
%
Note that for a small $J$ the suppression coefficient $C\sim J$ and, hence, close to the zeros of the Bessel function the spreading rate scales as 
%**********************************************************
\begin{equation}
\label{4d}
\frac{d \sigma}{dt} \sim {\cal J}_1^2\Big({\textstyle\frac{F_\omega}{F}}\Big) \;,
\end{equation}
which should be opposed to the scaling $d\sigma/dt\sim |{\cal J}_1(F_\omega/F)|$ for vanishing interactions. This effect is illustrated in the lower panel in Fig.~\ref{fig18}, showing the width $\sigma(t)$ at finite time $t=200\pi$ in dependence on the magnitude of the driving force. The two broken curves in the figure are results of numerical simulations of the system dynamics for $g=10$ and $=40$, and the dashed red curve reproduces the dependence (\ref{4b}) with $J$ substituted by $\widetilde{J}$. A qualitative agreement with experimental results \cite{Sias08,communic} is noticed.
%###########################
%###########################
\begin{figure}
\center
\includegraphics[width=8.5cm]{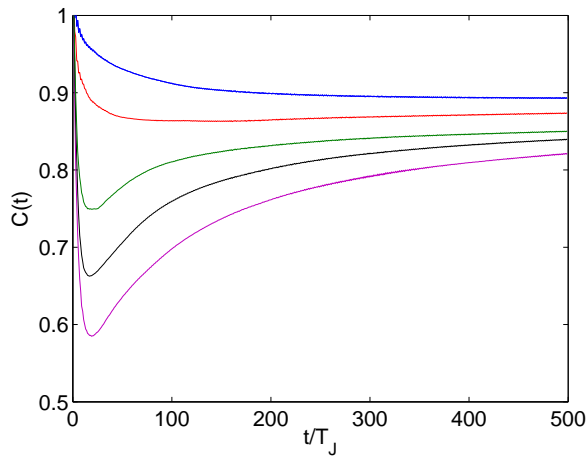}
\caption{Suppression coefficient $C(t)=\sigma(t;g\ne0)/\sigma(t;g=0)$ as functions of time for $J=2$, $dF=0.5$, $F_\omega=1.21F$, and interaction constant $g=5,10,20,30,40$ (from top to bottom). Incoherent initial conditions.}
\label{fig}
\end{figure}

For a quantitative comparison with experimental data a more detailed analysis  of the interaction-induced  suppression of ballistic spreading is needed. Figure \ref{fig} shows the suppression coefficient $C(t)=\sigma(t;g\ne0)/\sigma(t;g=0)$ as a function of time for $\widetilde{J}=1$, incoherent initial conditions, and different values of the interaction constant $g$. It is seen that for short times the spreading of interacting atoms is essentially suppressed as compared to the interaction-free case. Thus one may speak about temporal self-trapping. For long times this effect of interactions vanishes and the expansion regime becomes ballistic, i.e., $\sigma(t)\sim t$. These conclusions are consistent with the result depicted in the right-upper panel in Fig.~\ref{fig9}, showing characteristic density profile for $g\ne0$. Loosely speaking this density profile resembles a melting piece of ice, where the spreading becomes ballistic when ice melts completely.

%%%%%%%%%%%%%%%%%%%%%%%%%%%%%%%%%%%%
\section{Conclusion}
\label{sec4}

In conclusion, we have studied the dynamics of interacting cold atoms in a driven tilted optical lattice (\ref{1}).  Using the mean-field approach of the discrete nonlinear Schr\"odinger equation we were able to reproduce results of laboratory experiments \cite{Sias08,Hall10}, which report a resonant response of the system at driving frequencies $\omega\approx\omega_B/n$, where $\omega_B=h/dF$ is the Bloch frequency. 

Our contribution in understanding the experimental findings is as follows. It is shown that  the atomic dynamics in the vicinity of any of the resonances (in this paper we focused on the main resonance $\omega\approx\omega_B$) can be described in terms of the effective Hamiltonian (\ref{11a}), which formally corresponds to a stationary lattice with an effective static force $d\widetilde{F}=\hbar(\omega-\omega_B)$ and renormalized hopping matrix element $\widetilde{J}=J{\cal J}_1(F_\omega/F)$. Additionally, one has to substitute the initial coherent state, which corresponds to a BEC of atoms in a laboratory experiment, by an incoherent state.  The physics behind this seemingly artificial change of initial conditions is that, in a driven lattice, a BEC state rapidly decoheres in the presence of atom-atom interactions.

After reformulation of the problem in terms of an effective Hamiltonian, the atomic dynamics in a driven lattice can be mapped to that in a stationary lattice. Namely, for off-resonant driving one observes a subdiffusive spreading of the atomic cloud, where the cloud width grows $\sim t^{1/4}$. For resonant driving one meets the suppressed ballistic spreading (\ref{4d}), where the cloud width grows $\sim t$.

\section*{Acknowledgments}

Support from the Deutsche Forschungsgemeinschaft via the Graduiertenkolleg  
`Nichtlineare Optik und Ultrakurzzeitphysik' is gratefully acknowledged.

%%%%%%%%%%%%%%%%%%%%%%%%%%


\begin{thebibliography}{10}

%%%%%%%%%%%x1
\bibitem{Mors06}
 O. Morsch and M. Oberthaler, 
 %{\em Bose-Einstein condensates in optical lattices},  
 Rev. Mod. Phys. {\bf 78}, 179 (2006). 

\bibitem{Jona03}
M. Jona-Lasinio, O. Morsch, M. Cristiani, N. Malossi, J. H. M\"uller, E. Courtade, M. Anderlini,  and E. Arimondo,  
%{\em Asymmetric Landau-Zener tunneling in a periodic potential}, 
Phys. Rev. Lett. {\bf 91}, 230406 (2003).

%%%%%%%%%%%x2
\bibitem{Mors01}
O. Morsch, J. H. M\"uller, M. Cristiani, D. Ciampini, and E. Arimondo,  
%{\it  Bloch oscillations and mean-field effects of {Bose-Einstein} condensates in 1D optical lattices},  
Phys. Rev. Lett.  {\bf 87}, 140402 (2001).

\bibitem{Fatt08}
M. Fattori, C. D'Errico, G. Roati, M. Zaccanti, M. Jona-Lasinio, M. Modugno, M. Inguscio, and G. Modugno,  
%{\it Atom interferometry with a weakly  interacting {Bose-Einstein} condensate},  
Phys. Rev. Lett.  {\bf 100}, 080405 (2008).

\bibitem{Zhen04}
Yi. Zheng, M. Kostrun, and J. Javanainen,  
%{\it Low-acceleration instability of  a {Bose-Einstein} condensate in an optical lattice},  
Phys. Rev. Lett.  {\bf  93}, 230401 (2004).
  
\bibitem{Gust08a}
M. Gustavsson, E. Haller, M. J. Mark, J.  G. Danzl, G. Rojas-Kopeinig, and H. C. N\"{a}gerl, 
%{\it Control of interaction-induced dephasing of Bloch oscillations},  
Phys. Rev. Lett.  {\bf 100}, 080404 (2008).  

\bibitem{09BOBECa}
A. R. Kolovsky, H. J. Korsch and E. M. Graefe,
%{\em Bloch oscillations of Bose-Einstein condensates: Quantum counterpart of dynamical instability},
Phys. Rev. A {\bf 60}, 023617 (2009).


%%%%%%%%%%%%%x3
\bibitem{Bill08}
J. Billy, V. Josse, Z. Zuo, A. Bernard, B. Hambrecht, P. Lugan, D. Clement, L. Sanchez-Palencia, Ph. Bouyer, and A. Aspect, 
%{\em Direct observation of Anderson localization of matter-waves in a controlled disorder}, 
Nature {\bf 453}, 893 (2008).

\bibitem{Roat08}
G. Roati, C. D' Errico, L. Fallani, M. Fattori, C. Fort, M. Zaccanti, G. Modugno, M. Modugno, and M. Inguscio, 
%{\em Anderson localization of a non-interacting Bose-Einstein condensate},  
Nature {\bf 453}, 891 (2008).

\bibitem{Kopi08}
G. Kopidakis, S. Komineas, S. Flach, and S. Aubry, 
%{\em Absence of wavepacket diffusion in disodered nonlinear systems},
Phys. Rev. Lett. {\bf 100}, 084103 (2008) 

\bibitem{Piko08}
A. S. Pikovsky and D. L. Shepelyansky, 
%{\em Destruction of Anderson localization by a weak nonlinearity},
Phys. Rev. Lett. {\bf 100}, 094101 (2008)

%\bibitem{Schm98}
%P. Schmitteckert, T. Schulze, C. Schuster, P. Schwab, and U. Eckern, 
%{\em Anderson localization versus delocalization of interacting fermions in one dimension},
%Phys. Rev. Lett. {\bf 80}, 560 (1998) 

%\bibitem{Ogan07}
%V.~Oganesyan and D.~A.~Huse, 
%%Phys. Rev. B {\bf 75}, 155111 (2007)


%%%%%%%%%%%%x4
\bibitem{Lign07}
H. Lignier, C. Sias, D. Ciampini, Y. Singh, A. Zenesini, O. Morsch, and E. Arimondo,
%{\em Dynamical Control of Matter-Wave Tunneling in Periodic Potentials}
Phys. Rev. Lett. {\bf 99}, 220403 (2007).

\bibitem{Sias08}
C. Sias, H. Lignier, Y. P. Singh, A. Zenesini, D. Ciampini, O. Morsch, and E. Arimondo,
%{\em  Observation of photon-assisted tunneling in optical lattices},
Phys. Rev. Lett. {\bf 100}, 040404 (2008).

\bibitem{Ecka09}
A. Eckardt, M. Holthaus, H. Lignier, A. Zenesini, D. Ciampini, O. Morsch, and E. Arimondo
%{\em Exploring dynamic localization with a Bose-Einstein condensate},
Phys. Rev. A {\bf 79}, 013611 (2009).

\bibitem{Dunl86}
 D. H. Dunlap and V. M. Kenkre, 
 Phys. Rev. B {\bf 34}, 3625 (1986). 

\bibitem{Dres97}
K. Drese and M. Holthaus, 
%{\em Exploring a metal-insulator transition with ultracold atoms in standing light waves},
Phys. Rev. Lett. {\bf 78}, 2932 (1997).


%%%%%%%%%%%x5
\bibitem{preprint}
A. R. Kolovsky, E. A. G\'omez, and H. J. Korsch, 
%{\em Bose-Einstein condensates on tilted lattices: coherent, chaotic and subdiffusive dynamics},
Phys. Rev. A {\bf 81}, 025603 (2010).

\bibitem{Krim09}
D. O. Krimer, R. Khomeriki, and S. Flach, 
%{\em Delocalization and spreading in a nonlinear Stark ladder},
Phys. Rev. E {\bf 80}, 036201 (2009).

\bibitem{Gust08}
M. Gustavsson, E. Haller, M. J. Mark, J. G. Danzl, R. Hart, A. J. Daley, and H.-C. N\"agerl, 
%{\em Interference of interacting matter waves},
arXiv:0812.4836 (2008).

\bibitem{Hall10}
E. Haller, {\em et al.}, R.Hart, M.J.Mark, J.G.Danzl, L.Reichs\"ollner, and H.-Ch.N\"agerl, 
%{\em Inducing transport in a dissipation-free lattice with super Bloch oscillations}, 
Phys. Rev. Lett.  {\bf 104}, 200403 (2010).

\bibitem{communic}
E. Haller and H.-C. N\"agerl, private communication.


%%%%%%%%%%%%%%%x6
%\bibitem{PR}
%M. Gl\"uck, A. R. Kolovsky, and H. J. Korsch,
%{\em Wannier-Stark resonances in optical and semiconductor superlattices},
%Phys. Rep. {\bf 366} (2002)103, subsection 5.1.

\bibitem{Moss03}
H. J. Korsch and S. Mossmann,
%{\em An algebraic solution of driven single band tight binding dynamics},
Phys. Lett. A {\bf 317}, 54 (2003).

\bibitem{Thom04}
Q. Thommen, J.C.Garreau, and V.Zehnl\'e, 
%{\em Atomic motion in tilted optical lattices: An analytical approach}, 
J. Opt. B {\bf 6}, 301 (2004).

%\bibitem{comment} 
%A. R. Kolovsky and H. J. Korsch,
%{\em Comment to the paper by K.Drese and M.Holthaus},
%unpublished (1998).

\bibitem{remark1}
Another characteristic of the diffusion process is a so-called participation ration $L(t)=[\sum_l P_l^2(t)]^{-1}$. Since the distribution $P_l(t)$ shows no algebraic tails,  this quantity does not provide addition information and, roughly, $L(t)\sim \sigma(t)$.

\bibitem{remark2} 
To avoid a possible misunderstanding we recall once more that the chaotic regime assumes certain conditions for the system parameters \cite{Zhen04,09BOBECa}. Typically these conditions are satisfied for a weak static field.

%%%%%%%%%%%%%x7
\bibitem{Flach04} 
D. K. Campbell, S. Flach, and Y. S. Kivshar,
%{\em Localizing energy through nonlinearity and discreteness},
Phys. Today {\bf  57}, (2004) January, p. 43.

\bibitem{remark3}
Formation of a soliton state is a rather complicated process, sensitive to the initial state of the system. The mostly studied case corresponds to the population of a single well, where one has a reliable estimate for critical interactions, above which the discrete soliton is formed \cite{Moli92}. Unfortunately, we are not aware of any systematic analysis of the relevant to BEC dynamics case of a wide incoherent wave packet.

\bibitem{Moli92}
M. I. Molina and G. P. Tsironis, 
%{\em Dynamics of self-trapping in the discrete nonlinear Schr\"odinger equation}, 
Physica D: Nonlinear Phenomena {\bf 65}, 267 (1993).

\bibitem{remark4} 
Note that additional less pronounced peaks observed experimentally at $\omega/\omega_B=2,3/4,\ldots$ \cite{Hall10} are attributed to next-nearest neighbor hopping and can be described theoretically by a generalized single-band model \cite{Moss03}. 

%%%%%%%%%%%%x8
\bibitem{Albe09}
A. Alberti , V. V. Ivanov, G. M. Tino and G. Ferrari, 
%{\em Engineering the quantum transport of atomic wavefunctions over macroscopic distances},  
Nature Physics, {\bf 5}, 547 (2009).

%\bibitem{Ivan08}
%V.V.Ivanov, A.Alberti, M.Schioppo, G.Ferrari, M. Artoni, M.L.Chiofalo, and G.M.Tino, 
%{\em Coherent delocalization of atomic wave packets in driven lattice potentials}, 
%Phys. Rev. Lett. {\bf 100} (2008) 043602.

\bibitem{remark5}
The onset of dynamical instability for ordinary BO implies the interaction constant to be larger than some critical $g_{cr}$, which we identified as $g_{cr}\approx 2$.  Below $g_{cr}$ the system dynamics is quasi-regular and may correspond to the formation of stable bright solitons. This interesting regime of BEC dynamics will be discussed elsewhere.

\end{thebibliography}
\end{document}